# Multiphase $CO_2$ Dispersions in Microfluidics: Formation, Phases, and Mass Transfer

Tsai-Hsing Martin Ho, Dan Sameoto and Peichun Amy Tsai*

*Department of Mechanical Engineering, University of Alberta, Edmonton, Alberta T6G 2G8, Canada*

ABSTRACT

The dissolution and microfluidic mass transfer of carbon dioxide in water at high-pressure conditions are crucial for a myriad of technological applications, including microreactors, extractions, and carbon capture, utilization, and sequestration (CCUS) processes. In this experimental work, we use a high-pressure microfluidic method to elucidate the mass transfer process of $CO_2$ in water at high pressure. An intriguing multiphase $CO_2$ flow and dispersions are observed when operating at the pressure-temperature ($P$-$T$) condition close to the $CO_2$ gas-liquid phase boundary ($P = 6.5$ MPa and $T = 23.5 \pm 0.5$ °C). We propose a series of strategies to unravel this complex multi-phase dynamics by calculating each phase's volume and mass change in a gas-liquid coexistent $CO_2$ dispersion, estimating the possible $CO_2$ concentration change in water, and comparing with the $CO_2$ solubility data. Finally, we quantify the $CO_2$ mass transfer by directly calculating the $CO_2$ dissolution rate in water and estimating the volumetric mass transfer coefficient ($k_L a$). The results show that the mass transfer may be influenced by the specific area ($a$), $CO_2$ concentration gradient in the water slug, and the traveling speed of a dispersion.

## 1. Introduction

The technologies of carbon capture, utilization, and sequestration (CCUS) are essential to reduce $CO_2$ emissions and consequently mitigate global warming issues (Orr et al., 2018; Zhang et al., 2020). CCUS integrates methods to gather and transport $CO_2$ from large emission sites, such as power plants and oil refineries (Wilberforce et al., 2019). The concentrated $CO_2$ can be utilized for enhanced oil recovery (EOR) (Al Adasani and Bai, 2011) and enhanced coalbed methane (Prabu and Mallick, 2015; Mukherjee and Misra, 2018) for energy fuel production while sequestrating $CO_2$, and the further purified $CO_2$ can be used for chemical synthesis (Meunier et al., 2020; Branco et al., 2020). Alternatively, the captured $CO_2$ can be stored in various geological formations, such as in ocean (Adams and Caldeira, 2008; Aminu et al., 2017), in sub-seabeds (Carroll et al., 2014; Teng and Zhang, 2018), and in deep saline aquifers for a long time (Bachu, 2015).

The $CO_2$ mass transport behavior, therefore, plays a critical role in many CCUS processes, including the $CO_2$ absorption in the capture process (Abolhasani et al., 2014; Wilberforce et al., 2019) and the $CO_2$ dissolution in the pore fluids of sequestration sites. For example, in EOR engineering precise data of $CO_2$ dissolution in the pore-fluids helps estimate the required injection pressure to reach the minimum miscibility pressure for a better recovery efficiency (Al Adasani and Bai, 2011). The $CO_2$ dissolution and mass transfer processes affect the storage capacity of deep saline aquifers during the early injection stage (via structural and residual trapping of $CO_2$ in micropores) as well as the long-term solubility trapping and mineralization of $CO_2$ with the host, porous rock (Huppert and Neufeld, 2014; Bachu, 2015; Aminu et al., 2017). Furthermore, understanding the $CO_2$ mass transport processes helps evaluate the sealing integrity and calculate the pressure build-up and $CO_2$ leakage (Singh, 2018), which are essential for mitigating potential environmental impacts (Carroll et al., 2014; Blackford et al., 2014).

The early experimental investigations for measuring the $CO_2$ dissolution rate and mass transfer include both in-situ field experiments (Brewer et al., 1999, 2002; Blackford et al., 2014; Sellami et al., 2015) and laboratory measurements using stirred vessels (Van't Riet, 1979) or bubble column reactors (Shah et al., 1982; Kantarci et al., 2005). However, the field experiments usually are time- and budget-consuming, and most of the previous laboratory experiments require large apparatuses to handle working fluids at high pressure safely (Teng and Yamasaki, 2000; Ogasawara et al., 2001; Maalej et al., 2001) and therefore more space, resources, and experimental time. Hence, it motivates us to explore alternatives for acquiring essential mass transfer data at elevated pressure more efficiently since the relevant experimental studies are crucial but rare currently.

Microfluidics has become an emerging and useful tool for fluid physics (Anna, 2016; Amstad et al., 2017) as well as energy and environmental technologies due to its time and economic efficiency (De Jong et al., 2006; Kjeang et al., 2009; Sinton, 2014; Lifton, 2016). A variety of microfluidic platforms have been utilized to investigate the mass transfer rate of gases (e.g., $CO_2$ (Yue et al., 2007; Sun and Cubaud, 2011; Abolhasani et al., 2012), air (Yue et al., 2009), and ozone (Ren et al., 2012)) in different solvents, the $CO_2$ take-up capacity of physical solvents for $CO_2$ capture applications (Sun and Cubaud, 2011; Lefortier et al., 2012; Abolhasani et al., 2012), and the influence of temperature (Tumarkin et al., 2011) and surfactants (Shim et al., 2014) on $CO_2$ solubility in water. Moreover, the development of high-pressure microfluidic devices in the last decade enables ex-periments to operate in a broader pressure ($P$) and temperature ($T$) range up to $P = 45$ MPa and $T = 500$ °C (Marre et al., 2012). A few of microfluidic studies exploiting high-$P$ platforms provide insights into the $CO_2$ behaviors in different carbon storage scenarios, including the invasion patterns of $CO_2$ displacing the pore fluid (Zhang et al., 2011; Morais et al., 2016), the applications in the enhanced oil recovery (EOR) (Nguyen et al., 2015; Sharbatian et al., 2018), the physical properties at the supercritical state (Pinho et al., 2015), fast screening the $CO_2$ phase state in different solvents (Pinho et al., 2014; Bao et al., 2016), and the $CO_2$ solubility in brine (Liu et al., 2012).

*Corresponding author
E-mail address: peichun.amy.tsai@ualberta.ca
ORCID(s): 0000-0002-3095-3991 (P.A. Tsai)

Microfluidic studies addressing the mass transfer rate of $CO_2$ in water/brine at high $P$-$T$ conditions, especially for the liquid and supercritical $CO_2$, are still limited (Sell et al., 2013; Yao et al., 2015, 2017; Qin et al., 2017, 2018). Sell et al. (Sell et al., 2013) reported a minor influence of pressure on the $CO_2$ diffusion coefficient as $P$ increased from 0.1 to 5 Mpa. In contrast, the salinity significantly hinders $CO_2$ diffusion process in water, with a decrease in the diffusion coefficient by 60% with increasing salinity (from 0 to 5 M).

Yao et al. studied the pressure effects on the $CO_2$ mass transfer rate in water by measuring the liquid side volumetric mass transfer coefficient, $k_L a$, under elevated pressure from 0.1 to 3 M Pa. They found that $k_L a$ increased with the rising pressure and attributed this to the enlarged cross-sectional area of a $CO_2$ bubble at high pressure, which increases the $CO_2$-water contact area and enhances $CO_2$ mass transfer (Yao et al., 2015). Interestingly, they did not observe a significant influence of pressure on the $CO_2$ absorption in the chemical solvent DEA (diethanolamine) from their later experiments. They attributed to the shrinking interfacial area due to the altering flow patterns and the channel geometry (Yao et al., 2017).

Qin et al. calculated the mass transfer coefficient ($k_L$) from the shrinkage of supercritical $CO_2$ at 8 MPa and 40 °C according to the 3-D morphology of an ideal Taylor bubble in a rectangular channel (Qin et al., 2017). Their results showed that $k_L$ rose from $1.5 \times 10^{-4}$ to $7.5 \times 10^{-4}$ m/s as the water volume fraction increased from 0.28 to 0.9. Additionally, $CO_2$ droplets with a faster-moving speed had a high $k_L$ due to the strong inner recirculation in the water slugs, enhancing the mixing of $CO_2$ and water (Qin et al., 2018).

In this work, we experimentally investigate the dissolution of liquid $CO_2$ in water at the injection pressure $P_{inj} = 6.5$ MPa and room temperature, $T = 23.5$ °C. To the best knowledge of ours, we observe intriguing and remarkable multiphase-coexisting $CO_2$ dispersions at high pressure, for the first time, generated by a two-phase $CO_2$ flow merging with water at the microfluidic T-junction (see Figure 1), while most similar microfluidic studies using two-phase flow have focused on the generation of monodisperse segmented flows and associated fluid dynamics and patterns (Zhu et al., 2016; Laborie et al., 2016; Eggersdorfer et al., 2018; Chakraborty et al., 2019; Salari et al., 2020; Fan et al., 2020). To analyze the complex dynamics of the multi-component $CO_2$ droplets, which involves phase change and dissolution processes, we first differentiate the phase state of $CO_2$ inside the dispersion by comparing the mass change and the solubility limit in water (for a particular $P$ and $T$ condition). A detailed discussion is provided according to two different phase combinations. Based on these results, we further quantify the mass transfer process by calculating the dissolution rate of $CO_2$, $\dot{M}_{dis}$, and calculating the volumetric mass transfer coefficient, $k_L a$. We find that the estimated $k_L a$ decreases rapidly with time, which may attribute to the change of the specific area, $CO_2$ concentration in the water slug, and the moving speed of dispersions.

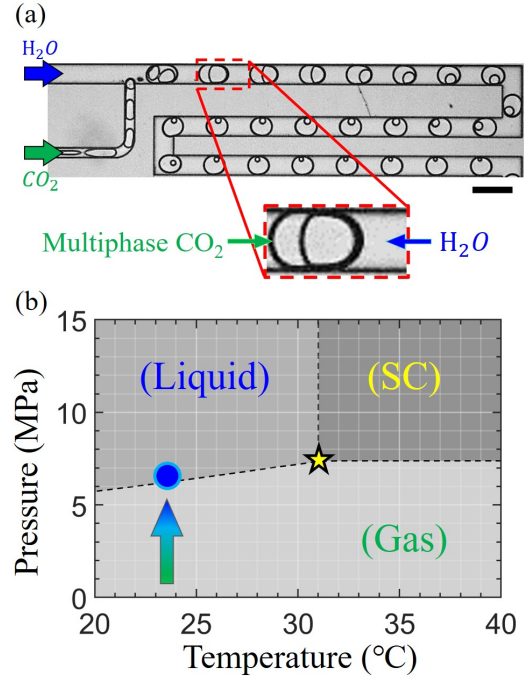

Figure 1: (a) An experimental snapshot of dispersions, containing multiphase (gas/liquid) $CO_2$ surrounded by the continuous phase of water, generated with a microfluidic T-junction. The scale bar represents 200 $\mu$m. (b) Plotted in the $CO_2$ phase diagram are the experimental pressure and temperature conditions (shown by ●) where the coexisting multiphase dispersions are observed. Experimentally, the pressure of $CO_2$ was gradually increased to $\approx 6.5$ MPa (●) at room temperature, as the arrow indicates. The $CO_2$ saturation pressure at 23.5°C is 6.14 MPa, as the black dashed line shows the gas-liquid phase boundary of $CO_2$. The phase diagram is plotted based on the database provided by the National Institute of Standards and Technology (NIST) (NIST, 2020).

## 2. Material and methods

We experimentally observed an intriguing two-phase $CO_2$ flow with a clear interface in the $CO_2$ injection channel before the microfluidic T-junction. Consequently, the multiphase coexisting $CO_2$ dispersions were formed in the T-junction microchannel after merging with Milli-Q water, as shown in Fig. 1a. The experiments were operated at the in-jection pressure, $P_{inj} = 6.5 \pm 0.05$ MPa and the chip temperature, $T_{chip} = 23.5 \pm 0.5$ °C, as pointed by the blue circle (●) in Fig. 1b. The microchannel was fabricated using the deep reactive ion etching method (DRIE) (Franssila, 2010; Ho and Tsai, 2020). The main channel is 100 $\mu$m in width and 30 $\mu$m in height. The side-channel for introducing $CO_2$ is 50 $\mu$m in width and has the same height as the main channel.

## 2.1. Experimental

The microfluidics was installed in a metal platform that can sustain high-pressure. $CO_2$ pressure was ramped up from the atmospheric condition (0.1 MPa and 23.5 °C) to a high-pressure condition for transforming to the liquid state (6.5 MPa and 23.5 °C). This process was controlled by a high-pressure gas pump (ISCO 100DX), which directly connects with the gas tank (Praxair, RES K $CO_2$ 99.998%). A back-pressure regulator (TESCOM BP 25 − 4000 PSI) was connected to the system's outlet to control the $CO_2$ flow rate by regulating a proper pressure gradient across the channel. Milli-Q water was loaded in stainless steel syringes and pumped by a syringe pump (Chemyx Inc. Fusion 6000) after one hour of degassing in a vacuum chamber. Water is injected at the flow rate of 10 $\mu$l/min for the bubbly, intermittent, and annular-1 flow. The water injection rate for the annular-2 flow is lower, at 5 $\mu$l/min. A thermocouple (K-type) was attached to the microfluidic chip and to monitor the temperature. The flow was observed by using an inverted microscope (Zeiss Axio Observer 7 Materials, with a 5× objective) and recorded by a high-speed camera (Phantom V710L) at 5,000 frames per second (fps).

## 2.2. Image analysis

We applied a series of post-image processing functions in ImageJ (NIH Image) (Abràmoff et al., 2004) to measure the size of a single dispersion and track its position with respect to time. We first calibrated these data by comparing them with the manually measured results. The maximum difference between them is within 2.3 % in length. The measured data were further analyzed by using a customized code written in MATLAB (MathWorks ®). The variation between dispersions was evaluated by calculating the standard deviation of at least five bubbles for each representative flow pattern. The result showed good consistency in the size and position. The variance is about 2.5 % maximum in dispersion's length and within 2.9 % in displacement.

## 2.3. Estimations of the total volume and surface area of a $CO_2$ dispersion

From our experimental images we notice that a dispersion comprises a brighter area surrounded by the back contour, as shown in Fig. 2a. We consider this black contour line resulting from the curved $CO_2$-water interface, which deflects the light and reduces the interface's brightness. The width of the black contour is measured to be about 8 $\mu$m. Therefore, we assume the geometry of a $CO_2$ dispersion is a planar disk, with two spherical caps on both ends (see

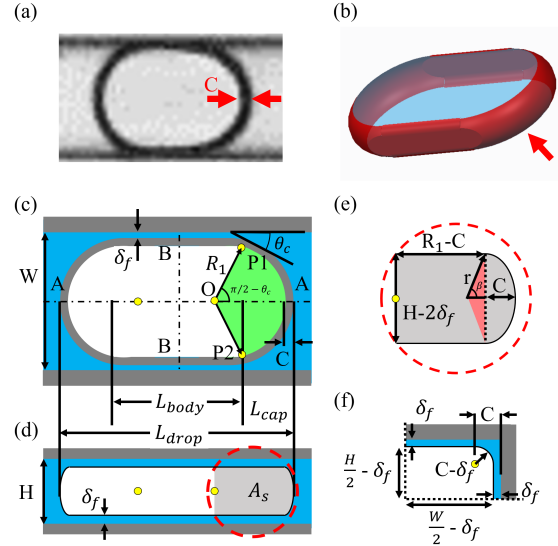

**Figure 2:** The schematic of an idealized microfluidic droplet, whose volume is comprised of a central section and two spherical caps on both ends. (a) Experimental snapshot of a $CO_2$ droplet. The black contour has a constant width, $C$ ($\approx 8\ \mu m$), for each droplet. (b) Schematic 3D model of $CO_2$ droplet volume. The red area is the curved $CO_2$-water interface, resulting in the shadow area shown in (a). (c) The top view of a droplet with a total length of $L_{drop} = 2L_{cap} + L_{body}$. $L_{cap}$ is the length of the circular caps, and $L_{body}$ denotes the central section of a $CO_2$ droplet, which is assumed to have a liquid film of a constant thickness, $\delta_f$, between the drop and sidewall. $R_1$ represents the radius of spherical caps that contact the sidewall with a contact angle, $\theta_c$. $C$ is estimated as the width of the dark area surrounding a droplet shown in (a). Both $\theta_c$ and $C$ are measured from experiments. (d) The schematic drawing for the droplet's side-view. $A_s$ denotes the cross-sectional area of the fan region (OP1P2) highlighted in color green in (c). (e) The detailed dimensions of the gray area $A_s$ in (d). (f) The drawing of the quarter cross-section of the center body through the B-B plane.

Fig. 2b). The red area represents the curved $CO_2$-water interface. Fig. 2c and 2d show the schematic shape of the $CO_2$ dispersion from the top view and the side view, respectively. The volume of a dispersion, $V_T$, is the sum of the center planar part and the two-sided caps:

$$V_T = 2V_{cap} + V_{body}. \qquad (1)$$

The volume of the cap $V_{cap}$ equals the volume of the green fan minus the volume of the triangle OP1P2, from Fig. 2c. The volume of the green fan, $V_{fan}$, and triangle OP1P2, $V_{tri}$, can be written as:

$$V_{fan} = \frac{1}{2}\int_{-(\frac{\pi}{2}-\theta_c)}^{\frac{\pi}{2}-\theta_c} A_s R_1 d\theta, \qquad (2)$$

$$V_{tri} = \frac{1}{2}(W - 2\delta_f)R_1 \cos\left(\frac{\pi}{2} - \theta_c\right)(H - 2\delta_f), \qquad (3)$$

where $W$ and $H$ are the width and height of the channel, $R_1$ is the radius of the circular cap, $\theta_c$ represents the con-tact angle between the (outer) cap and the microchannel (see Fig. 2c), and $\delta_f$ denotes the thickness of water film between $CO_2$ dispersion and the sidewall. $A_s$ in Eqn. (2) is the cross-sectional area of the green fan through the A-A plane, as the gray area in Fig. 2d. The dimensions of $A_s$ can be seen in Fig. 2e; note that the dimensions presented are exaggerated for better demonstration. $A_S$ can be expressed as:

$$A_S = (R_1 - C)(H - 2\delta_f) + \beta r^2 - \frac{1}{2} r \cos \beta (H - 2\delta_f), \quad (4)$$

where $r$ is the radius of the circular cap projected from the side view, and $\beta$ is half of the central angle. We acquired these two parameters according to their geometrical relations:

$$r \cos \beta = r - C, \quad (5)$$

$$r \sin \beta = \frac{H}{2} - \delta_f. \quad (6)$$

The volume of the center body of a $CO_2$ dispersion is calculated as $V_{body} = A_c(L_{drop} - 2L_{cap})$, where $L_{cap} = R_1 \cos(\frac{\pi}{2} - \theta_c)$ is the length of the circular cap on both ends. $A_c$ denotes the cross-sectional area of the center body through the B-B plane. We assume the center section's cross-sectional area is constant. Fig. 2f schematically demonstrates the idealized dimensions of this cross-section. We consider the shaded region of the body section is due to the curvilinear $CO_2$-water interface. $A_c$ hence can be expressed as:

$$A_c = (W - 2\delta_f)(H - 2\delta_f) - (4 - \pi)(C - \delta_f)^2. \quad (7)$$

By substituting Eqn. (2), (3), and (7) into Eqn. (1), the volume of $CO_2$ dispersion is acquired. This calculation result agree well with the 3-D model built using computer-aided design software (PTC Creo), with an estimated, average error of 1.5%. Furthermore, a close examination of the $CO_2$ bubble/droplet shows a slightly asymmetric shape of 1 $\mu$m difference between the front and rear radii. We calculate the corresponding $CO_2$ bubble's volume using Eqn. (1) to Eqn. (7). The volume difference due to the asymmetric bubble geometry is about 0.4%.

## 3. Results and discussion

### 3.1. Multiphase $CO_2$ flow and dispersions

Fig. 3a–c are experimental snapshots, showing three distinct $CO_2$ multiphase flow patterns observed in the $CO_2$ channel before the T-junction (when $P_{inj} = 6.51$ and $6.48$ MPa, $P_{back} = 6.15$ MPa and $T_{chip} \approx 23.5°C$, close to the $CO_2$ gas−liquid phase boundary). The multiphase flow quickly transited from a segmented flow pattern (Fig. 3a) to a parallel flow with an upstream pinch-off nose, emitting segmented flow periodically (Fig. 3b). This parallel flow kept propagating along with the $CO_2$ channel and finally anchored at the T-junction, seen in (c). These three flow modes are similar to the classic condensation flows observed when flowing the vapor state of water (Wu and Cheng, 2005; Chen et al., 2014) or refrigerants (Coleman and Garimella, 2003; Al-Zaidi et al., 2018) in the microchannel as the sidewall temperature ($T_w$) is lower than the vapor temperature of the fluid. We hence follow the same naming convention, calling the first segmented flow the "Bubbly flow" (Fig. 3a), the "Intermittent flow" for the second transitional pattern (in Fig. 3b), and the "Annular flow" for the final stratified flow pattern (in Fig. 3c). We artificially dyed the central or inner dispersed phase in light green in the images for better visualization.

At the microfluidic T-junction, the multiphase $CO_2$ flow periodically generates and emits dispersions in the main channel after being sheared off by the water flow. Fig. 3d shows such image sequences of the development of $CO_2$ dispersions for the different $CO_2$ flow patterns shown in Fig. 3a-c. These dispersions are multiphase with a clear interfacial boundary observed between the phases. Although each of the multi-phases is known to be either gas or liquid $CO_2$ because of the experimental pressure condition, without presumptions each exact phase was not certain and needed to be determined first. From the sequential images in Fig. 3 a-c, the inner phase (artificially dyed in light green) prefers to stay on the right side of a $CO_2$ dispersion, while the outer phase accumulates on the left. For clarification and convenience, we call the phase on the right side of the $CO_2$ dispersion the "inner (dyed) phase" and the other the "outer (undyed) phase" based on their appearance before the microfluidic T-junction (see Fig. 3 a-c).

Fig. 3e shows the total length of $CO_2$ dispersions, $L_{drop}$, changing with traveling time. The size change of the inner (dyed) phase seems to influences $L_{drop}$ greatly. For dispersions generated by the bubbly flow (depicted by ●) and the intermittent flow (depicted by ■), $L_{drop}$ decreases due to the shrinkage of the inner (dyed) phase. $L_{drop}$ reaches the steady-state when the inner phase completely vanishes. In general, we observe the same behavior for most dispersions generated by the annular flow, as depicted by ◆ in Fig. 3d. However, we surprisingly observed a different size change behavior of inner (dyed) phase in some dispersions generated by the annular $CO_2$ flow. As shown in Fig. 3 (d4), the inner phase shrank at the beginning when the dispersion detached from the T-junction. However, it started to expand after the dispersion traveled around 60 ms in the main channel, leading to the growth of $CO_2$ dispersions after moving 55 mm away from the T-junction (depicted by ▲).

### 3.2. Determination of the phase states of a multiphase $CO_2$ dispersion

Before detailed analyses on $CO_2$ flow and dissolution dynamics, we have to first differentiate and analyze the exact phases in the multiphase $CO_2$ droplets/bubbles without any presumptions. Based on the experimental condition ($P_{inj} = 6.5$ MPa and $T_{chip} \approx 23.5$ °C), we knew the coexistence of liquid and gas $CO_2$ in a dispersion before the T-junction. However, it is challenging to identify the phase states from the experimental grayscale images directly. These phases' grayscale values were close, measured to be 133.7 (±5.8) and 132.3 (±6.3) for the inner (dyed) and outer (undyed) phases, respectively. In addition, the surface tension at the water-gas $CO_2$ interface ($\sigma_{W-G}$ = 31.7 mN/m) is very close to that at the water-liquid $CO_2$ ($\sigma_{W-L}$ = 30.8 mN/m) (Pereira et al., 2016). This implies that both gas and liquid $CO_2$ could affix to the water and wrap the other phase. Here, we discuss the processes determining the phase state of $CO_2$ dispersions in more detail by analyzing the volume change, density difference, mass loss, and solubility in water.

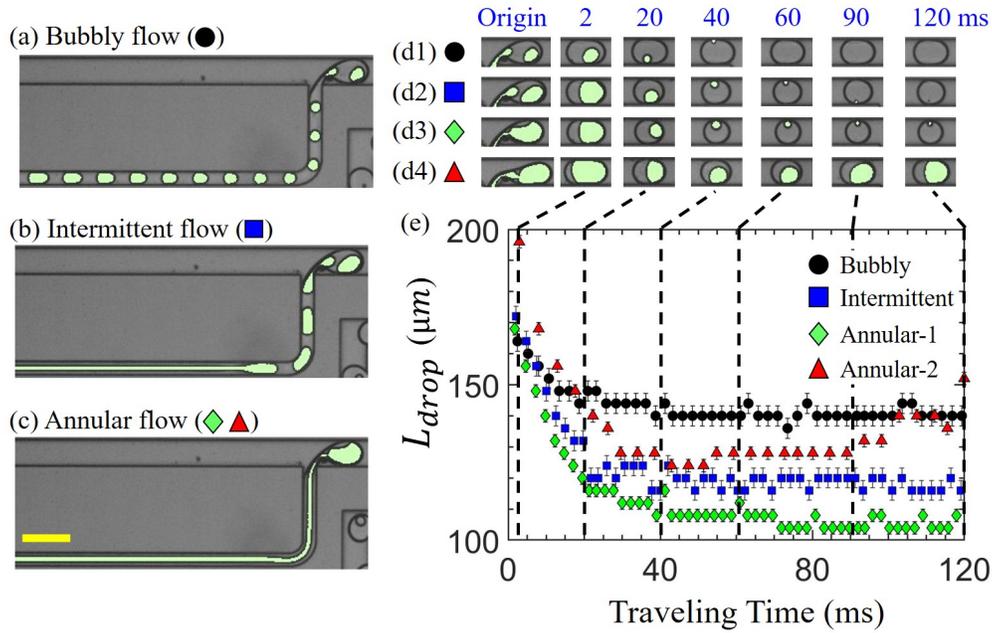

**Figure 3:** The snapshots of three different $CO_2$ flow patterns in our experiments: (a) the bubbly flow, (b) the intermittent flow, and (c) the annular flow. In the $CO_2$ channel, the inner part of $CO_2$ is highlighted in color light green. The yellow scale bar denotes 200 $\mu m$. (d) Representative image sequences of gas-liquid coexisting $CO_2$ dispersions in water of the bubbly flow in (d1), the intermittent flow in (d2), and two annular flow cases in (d3)-(d4) (multimedia view). (e) the total length, $L_{drop}$, of dispersions changes with time in the main channel after the microfluidic T-junction. Notably, we observed two different phase change modes in the $CO_2$ annular flow pattern. The first one is named Annular-1 (depicted by ◆), and the second one is called Annular-2 (depicted by ▲), respectively.

Fig. 4a demonstrates the total volume of $CO_2$ dispersions, $V_T$, changing with time, while 4b and 4c show the volume fraction of the inner (dyed) phase ($V_{in}$) and of outer (undyed) phase ($V_{out}$), respectively. From the observations, $V_{in}$ initially decreases dramatically for all representative cases, while $V_{out}$ increases in the range of 25% − 30% of the change of $V_{in}$, implying the mass exchange between two phases inside a $CO_2$ dispersion. Dispersions with a larger initial $V_{out}$ result in a bigger final size in the steady-state (flat regime), as seen in the bubbly flow (depicted by ●), comparing to other cases with larger $V_{in}$ (depicted by ■, ◆, and ▲).

To systematically determine each exact phase (gas or liquid) for the multiphase $CO_2$, we furthermore consider two different phase combinations and look into their corresponding mass change with time. By comparing with the $CO_2$ solubility data at 6.5 MPa and 23.5 °C, this data would help differentiate the phase state in a $CO_2$ dispersion. The first presumption considers that the inner (dyed) phase is liquid $CO_2$ and the outer gas. The second presumption is the other way, assuming that the inner (dyed) phase is gas and the outer liquid.

Fig. 5 shows the calculation results according to the two presumptions by showing the total cumulative mass, $M_T$, the net mass change, $\Delta M_T$, and the estimated $CO_2$ concentration in water slugs, $|\Delta M_T|/V_{slug}$. Under the first presumption (assuming the inner (dyed) phase is liquid $CO_2$), $M_T$ dramatically decreases in the first 20 ms and reaches the steady-state for dispersions of the bubbly, intermittent, and annular-1 flow, as shown in Fig. 5a. The net mass change, $\Delta M_T$, during this period is between $-5 \times 10^{-11}$ to $-20 \times 10^{-11}$ kg, shown in Fig. 5b. Notably, dispersions in Annular-2 (▲) show a gradual gain of the mass after 40 ms of traveling time, which is different from the previous three cases. Here, we consider that the mass transfer in a $CO_2$ dispersion includes two processes: (1) the phase transition between the two phases and (2) the dissolution of $CO_2$ into the surrounding water. The phase change processes should follow mass conservation, resulting in zero (or negligible) mass change (see Appendix A for a detailed analysis). Therefore, the net mass change, $\Delta M_T$, should always be negative due to the dissolution.

We further estimate the $CO_2$ concentration in the water changing with time by dividing the absolute value of $\Delta M$ by the volume of the adjacent water slug, $V_{slug}$. Fig. 5c shows that the estimated $CO_2$ concentration for all repre-sentative cases under presumption one is above 100 kg/m$^3$, much higher than the $CO_2$ solubility in water, $S = 61$ (kg/m$^3$), at the experimental conditions of $P_{inj} = 6.5$ MPa and $T_{chip} = 23.5$ °C (Diamond and Akinfiev, 2003). Fig. 5d–f reveal the analyzed $M_T$, $\Delta M_T$, and the estimated $CO_2$ concentration changing with time under the second presumption, respectively. The total cumulative mass of $CO_2$ dispersions decreases in a range of $1 \times 10^{-11} \sim 3 \times 10^{-11}$ kg, smaller than the results from the first presumption. All representative cases experienced an initial mass loss and reached steady-state at about 40 ms of traveling time. It should be noted that presumption two does not result in the strange mass gain in Annular-2 (depicted by ▲) in (d). Also, the net mass loss, $\Delta M_T = M_T - M_T(t = 0)$, is between $-1 \times 10^{-11}$ to $-3 \times 10^{-11}$ kg, giving the estimated $CO_2$ concentration in a range between 13.8 and 69.7 kg/m$^3$. This range is consistent with the $CO_2$ solubility data in the literature, $S = 61$ (kg/m$^3$) (Diamond and Akinfiev, 2003).

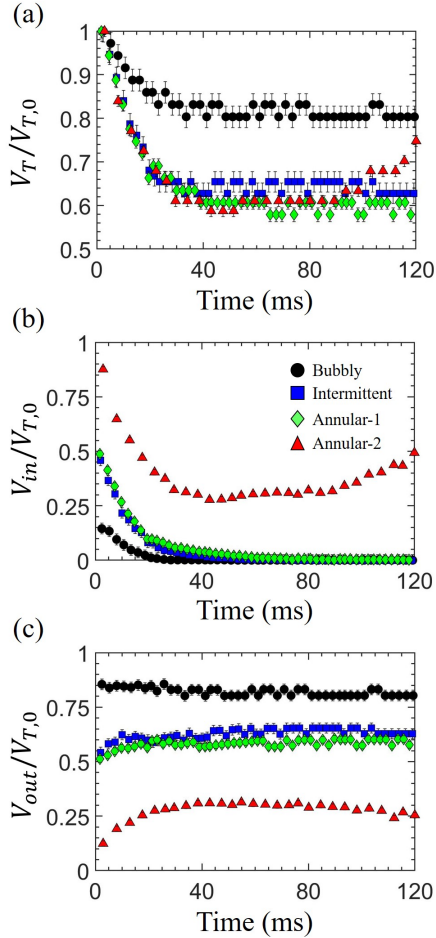

**Figure 4:** (a) The dimensionless volume of $CO_2$ dispersions changes with time, where $V_T$ is estimated according to the method introduced in Section 2.3 using Eq.(1)-(7), and $V_{T,0}$ denotes the initial volume of a dispersion, i.e., $V_T(t = 0)$. The corresponding fractional volume change of the inner (dyed) and outer phase inside the dispersion is presented in (b) and (c), respectively. dispersions generated by the three distinct $CO_2$ flow patterns are shown, including the bubbly flow (●), the intermittent flow (■), and the two different $CO_2$ phase change behavior observed in the annular flow [annular-1 (◆) and annular-2 (▲)].

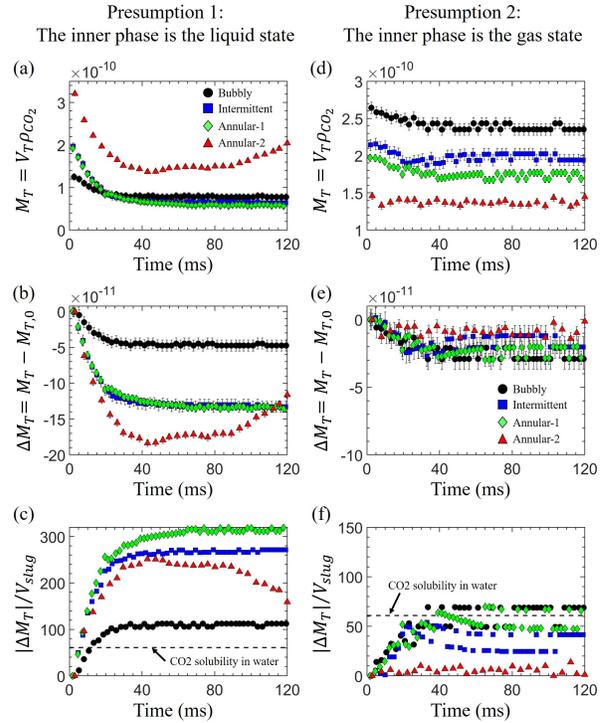

**Figure 5:** Results of time-varying $CO_2$ mass change estimated according to the presumption 1 and 2. (a) and (b) compare the cumulative mass of a $CO_2$ ispersion, $M_T = M_{in} + M_{out}$ (kg), changes with time, where $M_{in}$ and $M_{out}$ represent the mass of the inner (dyed) and outer phase of the dispersion, respectively. (c) and (d) show the net mass change of $CO_2$ dispersions, where $\Delta M_T = M_T - M_T(t = 0)$ (kg). Here, the negative sign of $\Delta M_T$ implies the dissolution of $CO_2$ in the water. (e) and (f) demonstrate the $CO_2$ concentration change in water, estimated by dividing the absolute value of $\Delta M_T$ by the volume of adjacent water slug, $V_{slug}$ (m$^3$). The black dash line denotes the theoretical solubility of $CO_2$ in water at $P = 6.5$ MPa and $T = 23.5$ °C (Diamond and Akinfiev, 2003).

The above analyses show that the second presumption yields consistent results with the $CO_2$ solubility data given $P$ and $T$. Therefore, we confirm that the inner (dyed) phase is $CO_2$ gas state, and the outer phase is the liquid state. This conclusion is further validated due to the following two reasons. First, the volume change ratio of the outer to inner phase, $V_{out}/V_{in}$, is about $0.25 - 0.3$. This value is also close to the density ratio of gas to liquid $CO_2$, $\rho_g/\rho_l = 0.33$, suggesting the shrinkage of the inner (dyed) phase in a $CO_2$ dispersion is the process combining the phase transition from the gas to liquid and the dissolution into the surrounding water. Second, presumption one yields an unusual increase in the total mass of case Annular-2 (demonstrated in Fig. 4a), implying that this assumption is incorrect because there are no other external mass sources. Besides, the presumption one overestimates the amount of dissolving $CO_2$ in water.

### 3.3. Formation of multiphase $CO_2$ dispersions

We further look into how the multiphase flow is formed in the $CO_2$ channel and what happens to such a $CO_2$ dispersion evolving in the microchannel, as seen in Fig. 3. These multiphase patterns are observed when increasing $P_{inj}$ from 1 atm up to 6.5 MPa at room temperature, and $CO_2$ in the pump started to phase change from the gas to liquid state slowly. Most of $CO_2$ likely stayed in the gas state as the oversaturated $CO_2$ vapor.

As $CO_2$ entered the microfluidics, $CO_2$ in the gas state flows in the center of the $CO_2$ channel, while the liquid $CO_2$ stayed on sidewalls because it is more viscous than the gas state. The formation of three distinct flow patterns observed might be similar to the jetting flow generated using the co-flow (Utada et al., 2007) and the cross-type microchannel (Cubaud and Mason, 2008). This kind of flow pattern may occur when the central fluid's velocity (gas $CO_2$) is greater than the outer one (liquid $CO_2$). The upstream annular flow may form because the inertia force from $CO_2$ gas stream stabilizes with the viscous force and surface tension at the gas-liquid interface (Utada et al., 2007). At the nose of the central $CO_2$ stream, gas bubbles detached and became the downstream bubbly flow, which might be triggered by the amplified capillary waves due to the surface tension gradient (Utada et al., 2008). The intermittent flow may be the transition between the bubbly and the annular flow. We observed that the intermittent flow emits droplets/bubbles downstream while upstream became an annular flow.

After the $CO_2$ flow merging with the continuous water phase, the multiphase $CO_2$ dispersions were formed periodically. The vapor $CO_2$ in a dispersion (i.e., the dyed inner phase) continued to transform from the gas to the liquid state, resulting in the rapid shrinkage of inner phase when moving in the main channel, as shown in Fig. 3d (first three rows). The expansion of dispersion in Annular-2 case (as demonstrated by ▲ in Fig. 3d and 3e) might due to the lower $CO_2$ vapor temperature than the microchannel's sidewall temperature. The local pressure along the channel keeps decreasing due to the hydraulic pressure gradient that drives the movement of a $CO_2$ dispersion (Kundu et al., 2011). $CO_2$ vapor temperature drops with the descending pressure, based on the Clausius-Clapeyron relation (Bejan, 2016), leading to the $P-T$ condition moving toward the phase boundary (Fig. 1b). Once the vapor temperature decreases to the sidewall temperature, $CO_2$ starts phase change from the liquid to gas state and expands $L_{drop}$.

### 3.4. Mass transfer of the multiphase $CO_2$ dispersions

To unravel the dissolution process while phase change is taking place, we develop an analysis to estimate the volumetric mass transfer coefficient, $k_L a$ (1/s), of $CO_2$ in water according to the dissolution rate of $CO_2$, $\dot{M}_{dis}$, and mass conservation. We first assume that the continuous phase, water, is initially free from $CO_2$. Both gas and liquid $CO_2$ are incompressible due to the negligible density variation in a small pressure difference (estimated ∼ 22 kPa). The $CO_2$ dissolution rate equals the net mass change rate of a dispersion, which can be expressed to the sum of the liquid and the gas parts:

$$\dot{M}_{dis} = \frac{dM_T}{dt} = \rho_l \frac{dV_l}{dt} + \rho_g \frac{dV_g}{dt}, \tag{8}$$

where $\rho_l$ ($\rho_g$) is the density of $CO_2$ in the liquid (gas) state. $V_l$ ($V_g$) denotes the volume of liquid (gas) $CO_2$. By integrating Eqn. (8), the total mass of dissolving $CO_2$ is expressed as:

$$\int_0^t \dot{M}_{dis} dt = \rho_l(V_l - V_{l,0}) + \rho_g(V_g - V_{g,0}), \tag{9}$$

where $V_{l,0}$ and $V_{g,0}$ (m$^3$) denotes the volume of liquid and gas $CO_2$ at $t = 0$, respectively.

The dissolved $CO_2$ increases its concentration in a water slug, and the convective mass transfer can be expressed as: (Yao et al., 2015; Ho et al., 2021)

$$\dot{M}_{dis} = -k_L A_T \left(C^* - C(t)\right) = -V_{slug} \frac{dC}{dt}. \tag{10}$$

Here, $k_L$ (m/s) denotes the mass transfer coefficient of $CO_2$ in water. $A_T$ (m$^2$) is the surface area of a dispersion, and $V_{slug}$ is the volume of the adjacent water slug. $C^*$ and $C(t)$ represent the saturation concentration of $CO_2$ in water and the mean $CO_2$ concentration in a water slug, respectively.

The $CO_2$ concentration change in the water slug, $C^* - C(t)$, can be written as an exponential function by integrating the second identity in Eqn. (10), so $C^* - C(t) = (C^* - C_0)e^{-k_L at} = Se^{-k_L at}$. The dissolved $CO_2$ in water as a function of time thus can be expressed as:

$$\dot{M}_{dis} = -k_L a S V_{slug} e^{-k_L at}, \tag{11}$$

where $C_0$ is the initial $CO_2$ concentration in water. Because we assume water is initially free from $CO_2$, $C^* - C_0$ represents the $CO_2$ solubility in the water ($S$) in the unit of kg/m$^3$. $a \equiv A_d/V_{slug}$; The volumetric mass transfer coefficient, $k_L a$, is associated with the mass change of $CO_2$ with time via integrating Eqn. (9) after substituting $\dot{M}_{dis}$ by Eqn. (11):

$$k_L a = -\frac{1}{t} \ln \left\{ 1 + \frac{1}{SV_{slug}} [\rho_l(V_l(t) - V_{l,0}) + \rho_g(V_g(t) - V_{g,0})] \right\}. \tag{12}$$

It should be noted that Eqn. (10) and Eqn. (11) consider a flat CO2-water interface for the mass transfer. Studies have shown that the Laplace pressure between the two phases can enhance gas molecules' transport of a stationary gas bubble into the surrounding solvent due to the overpressure. This enhancement becomes significant when the bubble radius is smaller than 15 $\mu$m and the solvent is nearly saturated (Kloek et al., 2001; Duncan and Needham, 2004). In our experiments, we degas water for an hour to ensure it is initially free from any dissolving gases, including $CO_2$. A large concentration difference between $CO_2$ dispersion and adjacent water would primarily drive the mass transfer of $CO_2$ (Duncan and Needham, 2004). In addition, the Laplace pressure at $P_{inj}$ = 6.5 MPa is estimated to be 1.5 kPa (about 0.023% of $P_{inj}$). Therefore, the influence of the Laplace pressure or surface curvature on the $CO_2$ mass transfer is likely negligible in our cases.

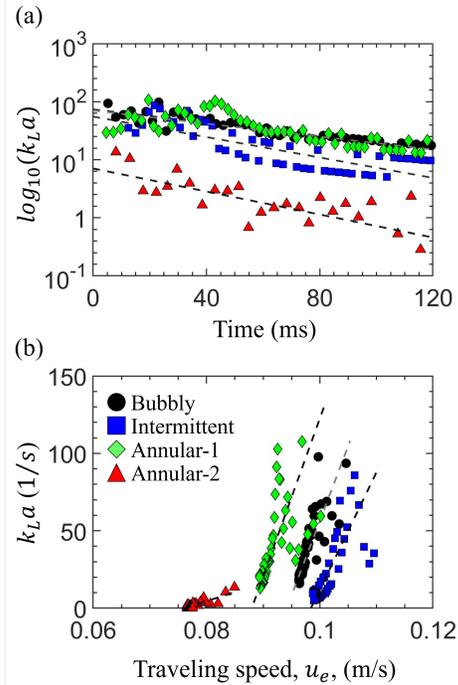

**Figure 6:** (a) The semi-log plot of estimated volumetric mass transfer coefficient, $k_L a$, of $CO_2$ changes with time. The black dashed lines are fitting curves showing the linear relation of $\log_{10}(k_L a)$ to time $t$. (b) The correlations of the estimated $k_L a$ and the traveling speed of $CO_2$ dispersions, $u_e$. The black dashed line represents the linear function of $k_L a = \alpha u_e + \beta$. Here, $u_e$ is defined by dividing the displacement from the origin by the traveling time.

Fig. 6a shows the semi-log plot of the estimated $k_L a$ for four representative cases by using Eqn. (12). All cases show a decreasing trend of $\log(k_L a)$ with a slop of $-7.5 \pm 2.2$. The black dashed line depicts the linear function as an eye guide. Overall, the average $k_L a$ is estimated to be 25 (1/s), which is consistent with our previous experimental results from injection pressures smaller than 8.5 MPa, giving the value of 29 (1/s) on average (Ho et al., 2021). This dynamic change in $k_L a$ was also observed and reported in a recent sub-millimeter scale experimental study of $CO_2$ mass transfer in water (Zhang et al., 2018).

The decreasing trend of $k_L a$ may attribute to the three following factors. First, the specific area ($a$) decreases with the reducing length of $CO_2$ dispersion $L_{drop}$ due to phase change and dissolution, where $a$ is defined by the ratio of the surface area of a $CO_2$ dispersion ($A_T$) and the sum of $V_T$ and $V_{slug}$. Second, the $CO_2$ concentration across the water-$CO_2$ interface is initially high as water contacts with $CO_2$ at $t \approx 0$, leading to the large $k_L a$ because $CO_2$ molecules dissolve in water in a very short time. The dissolution rate of $CO_2$ would quickly drop due to the surrounding water gradually saturated with $CO_2$. Finally, we found that the traveling speed of $CO_2$, $u_e$, influences $k_L a$ significantly.

Fig. 6b shows that $k_L a$ is the linear function of $\alpha \cdot u_e + \beta$ for all representative cases. The fitting coefficient $\alpha$ shows a larger value of 9150.9 ($\pm$1383.9) for the bubbly (●), intermittent (■), and annular-1 (◆) cases. A smaller $\alpha$ of 1162.9 is found for the annular-2 (▲). The faster $u_e$ allows fluid elements to pass the body part of $CO_2$ dispersions in a shorter time, preventing the fluid element from saturated with $CO_2$ and therefore increasing the mass transfer rate. Also, a high $u_e$ induces more intensive inner circulation inside the water slug, enhancing the mixing of dissolving $CO_2$ in the water (Günther et al., 2005; Kuhn and Jensen, 2012).

Furthermore, the annular-2 flow shows a smaller estimated $k_L a$ than the other three flow patterns. The data suggest that the pattern of the multiphase $CO_2$ flow and the water injection rate play a significant role in $CO_2$ mass transfer. The annular-2 flow is observed at the water injection rate of 5 $\mu$l/min, whereas the other cases are generated at a high water injection rate of 10 $\mu$l/min. The small water injection implies the weaker shear stress exerting from the continuous water on the dispersed $CO_2$ phase. As a result, the water phase takes a longer time to rupture the multiphase $CO_2$ stream and thus generates longer $CO_2$ dispersions (177.2 $\pm$2 $\mu$m) and water slugs (216.8$\pm$1.6 $\mu$m). The average length of $CO_2$ dispersions and water slugs for the bubbly, the intermittent, and the annular-1 flows are 162.9 $\pm$ 4 $\mu$m and 109.1 $\pm$ 4 $\mu$m, respectively. The long $CO_2$ dispersion may prohibit the $CO_2$ mass transfer due to the inactive liquid film surrounding the $CO_2$ dispersion (Irandoust et al., 1992; Pohorecki, 2007). In addition, a long water slug may also hinder the mass transfer of $CO_2$ to water due to the inefficient mixing inside the water slug (Bercic and Pintar, 1997; Yao et al., 2021). As a result, the annular-2 flow shows a small $k_L a$ than the other cases.

## 4. Conclusions

We experimentally investigate an intriguing dissolving dynamics of multiphase $CO_2$ dispersions in water at the P-T condition close to the gas-liquid phase boundary of $CO_2$ ($P$ = 6.5 MPa and $T$ = 23.5 $\pm$ 0.5 °C). The multiphase $CO_2$ dispersions are generated by merging the gas-liquid $CO_2$ flow with water using a microfluidic T-junction. Three flow patterns were observed sequentially, from the bubbly flow to the intermittent flow and the annular flow, representing the gaining dominance of the central $CO_2$ vapor stream over the liquid. A series of strategies are proposed and discussed to differentiate the phase state of $CO_2$ inside a dispersion, including estimating the volume of two distinct phases, calculating the corresponding mass change, and evaluating the possible $CO_2$ concentration change in the water. We find that the gas $CO_2$ with lower viscosity traveling faster at the center of the $CO_2$ channel and could be pitched off to form bubbles surrounded by liquid $CO_2$.

We further estimate the volumetric mass transfer coefficient, $k_L a$, from the multiphase dispersion dissolving into the surrounding water by considering both dissolution and the phase change processes. The estimated $k_L a$ in our experiments ($\approx 25$ s$^{-1}$) is greater than those from previous experimental data (of $0.3 < k_L a < 1.6$ s$^{-1}$) obtained when $P$ between 0.1 and 3 MPa (Yue et al., 2007, 2009; Ren et al., 2012; Yao et al., 2015). The larger $K_L a$ value found is likely attribute to the smaller hydraulic diameter used in our design ($d_c \approx 50$ $\mu$m), giving rise to a larger specific area (a) compared to previous work. Finally, the resultant $k_L a$ of the multiphase $CO_2$ dispersion shows an exponential decay with time due to decreasing $a$ and, moreover, a significantly linear increase with the $CO_2$ traveling speed, $u_e$.

## Conflict of Interest

There are no conflicts to declare.

## Acknowledgments

We gratefully acknowledge the support from Canada First Research Excellence Fund (CFREF), Future Energy System (FES T02-P05) at the University of Alberta, and Canada Foundation for Innovation (CFI). We also thank S. Bozic and S. Munro for their help with micro-fabrications at the nanoFAB in the University of Alberta. P.A.T holds a Canada Research Chair in Fluids and Interfaces and gratefully acknowledges funding from the Natural Sciences and Engineering Research Council of Canada (NSERC) and Alberta Innovates (AI). This research was undertaken, in part, thanks to funding from the Canada Research Chairs Program.

## Appendix

### A. Detailed calculations of mass change of each phase due to phase change and dissolution

This appendix shows the case bubbly flow's detailed mass change calculations for supporting the discussions regarding Fig. 5. Each representative case demonstrated in Fig. 5 has gone through this type of analysis for determining the phase state of a $CO_2$ dispersion. For a $CO_2$ dispersion flowing in the microchannel, the total volume and mass is the sum of the inner (dyed) and outer phase, as seen in Fig. 7a–c. Fig. 7b and 7c demonstrate the mass of a $CO_2$ dispersion varying with time based on the two different phase combinations, using presumption 1 and 2.

The total $CO_2$ volume and mass can be estimated from the contributions of the inner (dyed) and outer phases:

$$V_T(t) = V_{in}(t) + V_{out}(t), \quad (13)$$
$$M_T(t) = \rho_{CO_2} V_{in}(t) + \rho_{CO_2} V_{out}(t). \quad (14)$$

The mass change of each phase comes from two primary parts: due to the phase change and the dissolution in water (denoted by the subscript $P$ and $D$, respectively):

$$M_{out}(t) - M_{out}(t_0) = m_{out}^D + m_{out}^P, \quad (15)$$
$$M_{in}(t) - M_{in}(t_0) = m_{in}^D + m_{in}^P, \quad (16)$$

where $M_{out}$ ($M_{in}$) denotes the mass of the outer (dyed inner) phase. $m^P$ represents the mass change due to the phase transition between the outer and inner phases; $m^D$ is the mass loss due to its dissolution in water.

We consider the mass conservation during a phase change process and assume $m_{in}^P = -m_{out}^P$. The mass change of a $CO_2$ dispersion can be obtained by adding Eqn. (15) and Eqn. (16):

$$\Delta M_T(t) = M_T(t) - M_T(t_0) = m_{out}^D + m_{in}^D. \quad (17)$$

Fig. 7e–f present the results of $\Delta M_T(t)$ according to the two presumptions. Eqn. (17) implies that the mass of a $CO_2$ dispersion is always decreasing because $CO_2$ dissolves in water. Fig. 7d shows the estimated $CO_2$ concentration in the adjacent water slug, $C(t)$, compared with the theoretical $CO_2$ solubility in the water (black dashed-line). $C(t)$ was calculated by dividing the absolute value of $\Delta M_T$ in (e) and (f) by the volume of adjacent water slug, $V_{\text{slug}}$.

We can further estimate $m_{in}^P$ by subtracting Eqn. (16) from Eqn. (15):

$$m_{out}^D - m_{in}^D + 2m_{in}^P = \Delta M_{in}(t) - \Delta M_{out}(t). \quad (18)$$

Here, we assume the dissolved $CO_2$ from both phases is comparable. Thus, the term $m_{out}^D - m_{in}^D$ is negligible compared to the $2m_{in}^P$ term. $m_{in}^P$ can be expressed as half of the difference of $\Delta M_{in}(t)$ and $\Delta M_{out}(t)$:

$$m_{in}^P = \frac{1}{2} \left[ \Delta M_{in}(t) - \Delta M_{out}(t) \right]. \quad (19)$$

The results calculated from Eqn. (19) are shown in Fig. 7g.

## References

Abolhasani, M., Günther, A., Kumacheva, E., 2014. Microfluidic studies of carbon dioxide. Angew. Chem. Int. Ed. 53, 7992–8002.

Abolhasani, M., Singh, M., Kumacheva, E., Günther, A., 2012. Automated microfluidic platform for studies of carbon dioxide dissolution and solubility in physical solvents. Lab Chip 12, 1611–1618.

Abràmoff, M.D., Magalhães, P.J., Ram, S.J., 2004. Image processing with ImageJ. Biophotonics int. 11, 36–42.

Adams, E.E., Caldeira, K., 2008. Ocean storage of $CO_2$. Elements 4, 319–324.

Al Adasani, A., Bai, B., 2011. Analysis of EOR projects and updated screening criteria. J. Pet. Sci. Eng. 79, 10–24.

Al-Zaidi, A.H., Mahmoud, M.M., Karayiannis, T.G., 2018. Condensation flow patterns and heat transfer in horizontal microchannels. Exp. Therm Fluid Sci. 90, 153–173.

Aminu, M.D., Nabavi, S.A., Rochelle, C.A., Manovic, V., 2017. A review of developments in carbon dioxide storage. Appl. Energy 208, 1389–1419.

Amstad, E., Chen, X., Eggersdorfer, M., Cohen, N., Kodger, T.E., Ren, C.L., Weitz, D.A., 2017. Parallelization of microfluidic flow-focusing devices. Phys. Rev. E 95, 043105.

Anna, S.L., 2016. Droplets and bubbles in microfluidic devices. Annu. Rev. Fluid Mech. 48, 285–309.

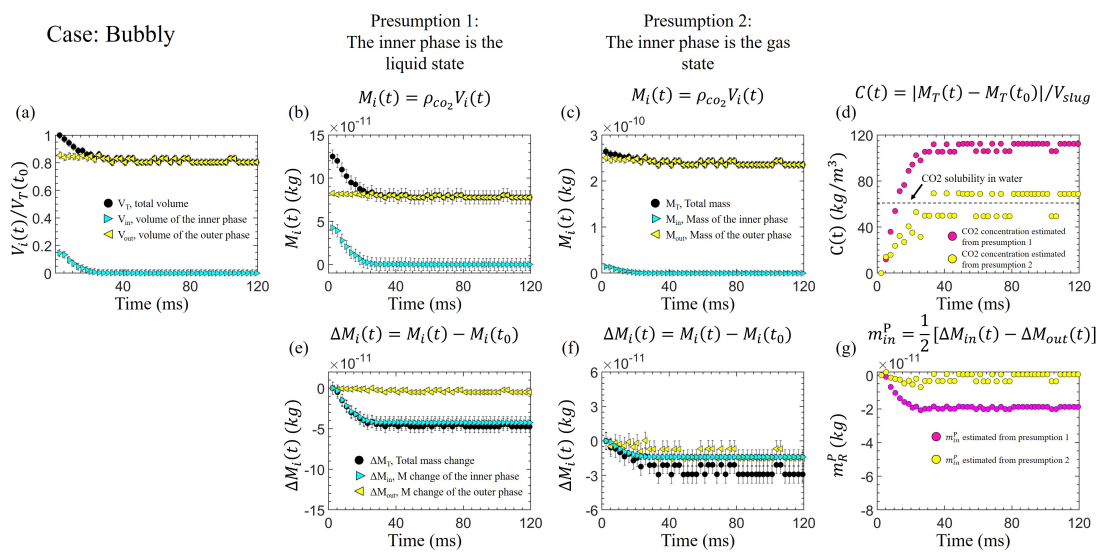

**Figure 7:** The detailed calculations of a $CO_2$ dispersion based on the mean value from five dispersions in the case of the bubbly flow. (a) The volume fractions, $V_i(t)$, of the outer phase ($V_{out}$), the (dyed) inner phase ($V_{in}$), and the total volume ($V_T = V_{out} + V_{in}$) changing with time ($t$). (b) The mass of the outer phase, $M_{out} = \rho_{CO_2} V_{out}$, and the dyed, inner phase, $M_{in} = \rho_{CO_2} V_{in}$, changing with time based on Presumption 1 (assuming the inner phase is the liquid $CO_2$). The total mass of a $CO_2$ dispersion, $M_T = M_{out} + M_{in}$. (c) Results of $M_{out}$, $M_{in}$, and $M_T$ calculated from Presumption 2 (assuming the outer phase is the liquid $CO_2$). (e) and (f) present the net mass change of a dispersion varying with time calculated according to presumption 1 and 2, respectively. $\Delta M$ represents the variation of mass from its initial value, $M(t_0)$. The data for the outer phase is denoted by ◁, while the dyed inner phase by ▷; ● represents the sum properties. (d) The estimated $CO_2$ concentration, $C(t)$, changes with time according to presumption 1 (●) and 2 (●). The black dashed line denotes the theoretical $CO_2$ solubility in the water. (g) The estimated mass change of the dyed, inner phase resulted from the phase transition.

Bachu, S., 2015. Review of $CO_2$ storage efficiency in deep saline aquifers. Int. J. Greenhouse Gas Control 40, 188–202.

Bao, B., Riordon, J., Xu, Y., Li, H., Sinton, D., 2016. Direct measurement of the fluid phase diagram. Anal. Chem. 88, 6986–6989.

Bejan, A., 2016. Advanced Engineering Thermodynamics. John Wiley & Sons.

Bercic, G., Pintar, A., 1997. The role of gas bubbles and liquid slug lengths on mass transport in the taylor flow through capillaries. Chem. Eng. Sci. 52, 3709–3719.

Blackford, J., Stahl, H., Bull, J.M., Bergès, B.J., Cevatoglu, M., Lichtschlag, A., Connelly, D., James, R.H., Kita, J., Long, D., et al., 2014. Detection and impacts of leakage from sub-seafloor deep geological carbon dioxide storage. Nat. Clim. Change 4, 1011–1016.

Branco, J.B., Brito, P.E., Ferreira, A.C., 2020. Methanation of $CO_2$ over nickel-lanthanide bimetallic oxides supported on silica. Chem. Eng. J. 380, 122465.

Brewer, P.G., Friederich, G., Peltzer, E.T., Orr, F.M., 1999. Direct experiments on the ocean disposal of fossil fuel $CO_2$. Science 284, 943–945.

Brewer, P.G., Peltzer, E.T., Friederich, G., Rehder, G., 2002. Experimental determination of the fate of rising $CO_2$ droplets in seawater. Environ. Sci. Technol. 36, 5441–5446.

Carroll, A., Przeslawski, R., Radke, L., Black, J., Picard, K., Moreau, J., Haese, R., Nichol, S., 2014. Environmental considerations for subseabed geological storage of $CO_2$: A review. Cont. Shelf Res. 83, 116–128.

Chakraborty, I., Ricouvier, J., Yazhgur, P., Tabeling, P., Leshansky, A., 2019. Droplet generation at hele-shaw microfluidic t-junction. Phys. Fluids 31, 022010.

Chen, Y., Shen, C., Shi, M., Peterson, G.P., 2014. Visualization study of flow condensation in hydrophobic microchannels. AIChE J. 60, 1182–1192.

Coleman, J.W., Garimella, S., 2003. Two-phase flow regimes in round, square and rectangular tubes during condensation of refrigerant R134a. Int. J. Refrig. 26, 117–128.

Cubaud, T., Mason, T.G., 2008. Capillary threads and viscous droplets in square microchannels. Phys. Fluids 20, 053302.

De Jong, J., Lammertink, R.G., Wessling, M., 2006. Membranes and microfluidics: a review. Lab Chip 6, 1125–1139.

Diamond, L.W., Akinfiev, N.N., 2003. Solubility of $CO_2$ in water from -1.5 to 100 °C and from 0.1 to 100 MPa: evaluation of literature data and thermodynamic modelling. Fluid Phase Equilib. 208, 265–290.

Duncan, P.B., Needham, D., 2004. Test of the epstein- plesset model for gas microparticle dissolution in aqueous media: effect of surface tension and gas undersaturation in solution. Langmuir 20, 2567–2578.

Eggersdorfer, M.L., Seybold, H., Ofner, A., Weitz, D.A., Studart, A.R., 2018. Wetting controls of droplet formation in step emulsification. PNAS 115, 9479–9484.

Fan, W.Y., Li, S.C., Li, L.X., Zhang, X., Du, M.Q., Yin, X.H., 2020. Hydrodynamics of gas/shear-thinning liquid two-phase flow in a co-flow mini-channel: Flow pattern and bubble length. Phys. Fluids 32, 092004.

Franssila, S., 2010. Introduction to Microfabrication. John Wiley & Sons.

Günther, A., Jhunjhunwala, M., Thalmann, M., Schmidt, M.A., Jensen, K.F., 2005. Micromixing of miscible liquids in segmented gas- liquid flow. Langmuir 21, 1547–1555.

Ho, T.H.M., Tsai, P.A., 2020. Microfluidic salt precipitation: implications for geological $CO_2$ storage. Lab Chip 20, 3806–3814.

Ho, T.H.M., Yang, J., Tsai, P.A., 2021. Microfluidic mass transfer of $CO_2$ at different phases. unpublished data .

Huppert, H.E., Neufeld, J.A., 2014. The fluid mechanics of carbon dioxide sequestration. Annu. Rev. Fluid Mech. 46, 255–272.

Irandoust, S., Ertlé, S., Andersson, B., 1992. Gas-liquid mass transfer in taylor flow through a capillary. Can. J. Chem. Eng. 70, 115–119.

Kantarci, N., Borak, F., Ulgen, K.O., 2005. Bubble column reactors. Process Biochem. 40, 2263–2283.

Kjeang, E., Djilali, N., Sinton, D., 2009. Microfluidic fuel cells: A review.

J. Power Sources 186, 353–369.

Kloek, W., Van Vliet, T., Meinders, M., 2001. Effect of bulk and interfacial rheological properties on bubble dissolution. J. Colloid Interface Sci. 237, 158–166.

Kuhn, S., Jensen, K.F., 2012. A ph-sensitive laser-induced fluorescence technique to monitor mass transfer in multiphase flows in microfluidic devices. Ind. Eng. Chem. Res. 51, 8999–9006.

Kundu, P.K., Cohen, I.M., Dowling, D.R., 2011. Fluid Mechanics, 5th ed. London: Academic.

Laborie, B., Rouyer, F., Angelescu, D., Lorenceau, E., 2016. On the stability of the production of bubbles in yield-stress fluid using flow-focusing and t-junction devices. Phys. Fluids 28, 063103.

Lefortier, S.G., Hamersma, P.J., Bardow, A., Kreutzer, M.T., 2012. Rapid microfluidic screening of $CO_2$ solubility and diffusion in pure and mixed solvents. Lab Chip 12, 3387–3391.

Lifton, V.A., 2016. Microfluidics: an enabling screening technology for enhanced oil recovery (EOR). Lab Chip 16, 1777–1796.

Liu, N., Aymonier, C., Lecoutre, C., Garrabos, Y., Marre, S., 2012. Microfluidic approach for studying $CO_2$ solubility in water and brine using confocal raman spectroscopy. Chem. Phys. Lett. 551, 139–143.

Maalej, S., Benadda, B., Otterbein, M., 2001. Influence of pressure on the hydrodynamics and mass transfer parameters of an agitated bubble reactor. Chem. Eng. Technol. 24, 77–84.

Marre, S., Roig, Y., Aymonier, C., 2012. Supercritical microfluidics: Opportunities in flow-through chemistry and materials science. J. Supercrit. Fluids 66, 251–264.

Meunier, N., Chauvy, R., Mouhoubi, S., Thomas, D., De Weireld, G., 2020. Alternative production of methanol from industrial $CO_2$. Renewable energy 146, 1192–1203.

Morais, S., Liu, N., Diouf, A., Bernard, D., Lecoutre, C., Garrabos, Y., Marre, S., 2016. Monitoring $CO_2$ invasion processes at the pore scale using geological labs on chip. Lab Chip 16, 3493–3502.

Mukherjee, M., Misra, S., 2018. A review of experimental research on enhanced coal bed methane (ECBM) recovery via $CO_2$ sequestration. Earth Sci. Rev. 179, 392–410.

Nguyen, P., Mohaddes, D., Riordon, J., Fadaei, H., Lele, P., Sinton, D., 2015. Fast fluorescence-based microfluidic method for measuring minimum miscibility pressure of $CO_2$ in crude oils. Anal. Chem. 87, 3160–3164.

NIST, 2020. NIST standard reference database number 69. URL: https://webbook.nist.gov/chemistry/.

Ogasawara, K., Yamasaki, A., Teng, H., 2001. Mass transfer from $CO_2$ drops traveling in high-pressure and low-temperature water. Energy Fuels 15, 147–150.

Orr, F.M., et al., 2018. Carbon capture, utilization, and storage: an update. SPE J. 23, 2–444.

Pereira, L.M., Chapoy, A., Burgass, R., Oliveira, M.B., Coutinho, J.A., Tohidi, B., 2016. Study of the impact of high temperatures and pressures on the equilibrium densities and interfacial tension of the carbon dioxide/water system. J. Chem. Thermodyn. 93, 404–415.

Pinho, B., Girardon, S., Bazer-Bachi, F., Bergeot, G., Marre, S., Aymonier, C., 2014. A microfluidic approach for investigating multicomponent system thermodynamics at high pressures and temperatures. Lab Chip 14, 3843–3849.

Pinho, B., Girardon, S., Bazer-Bachi, F., Bergeot, G., Marre, S., Aymonier, C., 2015. Simultaneous measurement of fluids density and viscosity using HP/HT capillary devices. J. Supercrit. Fluids 105, 186–192.

Pohorecki, R., 2007. Effectiveness of interfacial area for mass transfer in two-phase flow in microreactors. Chem. Eng. Sci. 62, 6495–6498.

Prabu, V., Mallick, N., 2015. Coalbed methane with $CO_2$ sequestration: An emerging clean coal technology in india. Renewable Sustainable Energy Rev. 50, 229–244.

Qin, N., Wen, J.Z., Chen, B., Ren, C.L., 2018. On nonequilibrium shrinkage of supercritical $CO_2$ droplets in a water-carrier microflow. Appl. Phys. Lett. 113, 033703.

Qin, N., Wen, J.Z., Ren, C.L., 2017. Highly pressurized partially miscible liquid-liquid flow in a micro-T-junction. I. experimental observations. Phys. Rev. E 95, 043110.

Ren, J., He, S., Ye, C., Chen, G., Sun, C., 2012. The ozone mass transfer characteristics and ozonation of pentachlorophenol in a novel microchannel reactor. Chem. Eng. J. 210, 374–384.

Salari, A., Xu, J., Kolios, M.C., Tsai, S.S., 2020. Expansion-mediated breakup of bubbles and droplets in microfluidics. Phys. Rev. Fluids 5, 013602.

Sell, A., Fadaei, H., Kim, M., Sinton, D., 2013. Measurement of $CO_2$ diffusivity for carbon sequestration: A microfluidic approach for reservoir-specific analysis. Environ. Sci. Technol. 47, 71–78.

Sellami, N., Dewar, M., Stahl, H., Chen, B., 2015. Dynamics of rising $CO_2$ bubble plumes in the qics field experiment: Part 1–the experiment. Int. J. Greenhouse Gas Control 38, 44–51.

Shah, Y., Kelkar, B.G., Godbole, S., Deckwer, W.D., 1982. Design parameters estimations for bubble column reactors. AIChE J. 28, 353–379.

Sharbatian, A., Abedini, A., Qi, Z., Sinton, D., 2018. Full characterization of $CO_2$–oil properties on-chip: solubility, diffusivity, extraction pressure, miscibility, and contact angle. Anal. Chem. 90, 2461–2467.

Shim, S., Wan, J., Hilgenfeldt, S., Panchal, P.D., Stone, H.A., 2014. Dissolution without disappearing: Multicomponent gas exchange for $CO_2$ bubbles in a microfluidic channel. Lab Chip 14, 2428–2436.

Singh, H., 2018. Impact of four different $CO_2$ injection schemes on extent of reservoir pressure and saturation. Adv. Geo-Energy Res. 2, 305–318.

Sinton, D., 2014. Energy: the microfluidic frontier. Lab Chip 14, 3127–3134.

Sun, R., Cubaud, T., 2011. Dissolution of carbon dioxide bubbles and microfluidic multiphase flows. Lab Chip 11, 2924–2928.

Teng, H., Yamasaki, A., 2000. Dissolution of buoyant $CO_2$ drops in a counterflow water tunnel simulating the deep ocean waters. Energy Convers. Manage. 41, 929–937.

Teng, Y., Zhang, D., 2018. Long-term viability of carbon sequestration in deep-sea sediments. Sci. Adv. 4, eaao6588.

Tumarkin, E., Nie, Z., Park, J.I., Abolhasani, M., Greener, J., Sherwood-Lollar, B., Günther, A., Kumacheva, E., 2011. Temperature-controlled 'breathing' of carbon dioxide bubbles. Lab Chip 11, 3545–3550.

Utada, A.S., Fernandez-Nieves, A., Gordillo, J.M., Weitz, D.A., 2008. Absolute instability of a liquid jet in a coflowing stream. Phys. Rev. Lett. 100, 014502.

Utada, A.S., Fernandez-Nieves, A., Stone, H.A., Weitz, D.A., 2007. Dripping to jetting transitions in coflowing liquid streams. Phys. Rev. Lett. 99, 094502.

Van't Riet, K., 1979. Review of measuring methods and results in nonviscous gas-liquid mass transfer in stirred vessels. Ind. Eng. Chem. Process Des. Dev. 18, 357–364.

Wilberforce, T., Baroutaji, A., Soudan, B., Al-Alami, A.H., Olabi, A.G., 2019. Outlook of carbon capture technology and challenges. Sci. Total Environ. 657, 56–72.

Wu, H., Cheng, P., 2005. Condensation flow patterns in silicon microchannels. Int. J. Heat Mass Transfer 48, 2186–2197.

Yao, C., Dong, Z., Zhao, Y., Chen, G., 2015. Gas-liquid flow and mass transfer in a microchannel under elevated pressures. Chem. Eng. Sci. 123, 137–145.

Yao, C., Zhao, Y., Ma, H., Liu, Y., Zhao, Q., Chen, G., 2021. Two-phase flow and mass transfer in microchannels: A review from local mechanism to global models. Chem. Eng. Sci. 229, 116017.

Yao, C., Zhu, K., Liu, Y., Liu, H., Jiao, F., Chen, G., 2017. Intensified $CO_2$ absorption in a microchannel reactor under elevated pressures. Chem. Eng. J. 319, 179–190.

Yue, J., Chen, G., Yuan, Q., Luo, L., Gonthier, Y., 2007. Hydrodynamics and mass transfer characteristics in gas–liquid flow through a rectangular microchannel. Chem. Eng. Sci. 62, 2096–2108.

Yue, J., Luo, L., Gonthier, Y., Chen, G., Yuan, Q., 2009. An experimental study of air–water taylor flow and mass transfer inside square microchannels. Chem. Eng. Sci. 64, 3697–3708.

Zhang, C., Oostrom, M., Grate, J.W., Wietsma, T.W., Warner, M.G., 2011. Liquid $CO_2$ displacement of water in a dual-permeability pore network micromodel. Environ. Sci. Technol. 45, 7581–7588.

Zhang, P., Yao, C., Ma, H., Jin, N., Zhang, X., Lü, H., Zhao, Y., 2018. Dynamic changes in gas-liquid mass transfer during taylor flow in long serpentine square microchannels. Chem. Eng. Sci. 182, 17–27.

Zhang, Z., Pan, S.Y., Li, H., Cai, J., Olabi, A.G., Anthony, E.J., Manovic, V., 2020. Recent advances in carbon dioxide utilization. Renewable Sustainable Energy Rev. 125, 109799.

Zhu, P., Kong, T., Lei, L., Tian, X., Kang, Z., Wang, L., 2016. Droplet breakup in expansion-contraction microchannels. Sci. Rep. 6, 1–11.